\documentclass{aa}
\usepackage{natbib}
\usepackage{rotfloat}
\usepackage{longtable}
\usepackage{graphicx}
\usepackage{relsize}
\usepackage{lscape}
\newcommand{\tol}[4][]{%
  \ifx f#1 \def\toltop{1.35ex}\def\tolbot{0ex}%
  \else \def\toltop{1ex}\def\tolbot{-0.35ex}\fi
  #2\hskip0.1em\rlap{\raisebox{\toltop}{$\mathsmaller{#3}$}}%
  \raisebox{\tolbot}{$\mathsmaller{#4}$}\hskip0.05em}

\bibpunct{(}{)}{;}{a}{}{,}

\begin{document}
\title{Spectroscopy of the neighboring massive clusters Abell~222 and
  Abell~223
  \thanks{Based on observations made at ESO -- La Silla (Chile)}}
\author{J. P. Dietrich\inst{1}
  \and 
  D. I. Clowe\inst{1}
  \and 
  G. Soucail\inst{2}}
\institute{Institut f\"ur Astrophysik und Extraterrestrische
  Forschung, Universit\"at Bonn, Auf dem H\"ugel 71, 53121, Bonn,
  Germany
  \and
  Observatoire Midi--Pyr\'en\'ees, UMR5572, 14 Av. \'Edouard Belin,
  31400 Toulouse, France}
\offprints{J.~P.~Dietrich,
  \email{dietrich@astro.uni-bonn.de}}
\date{Received 29.05.2002 / Accepted 09.08.2002}
\abstract{We present a spectroscopic catalog of the neighboring
  massive clusters \object{Abell~222} and \object{Abell~223}. The
  catalog contains the positions, redshifts, $R$ magnitudes, $V-R$
  color, as well as the equivalent widths for a number of lines for
  183 galaxies, 153 of them belonging to the A~222 and A~223 system.
  We determine the heliocentric redshifts to be $z=0.2126\pm0.0008$
  for A~222 and $z=0.2079\pm0.0008$ for A~223. The velocity
  dispersions of both clusters in the cluster restframe are about the
  same: $\sigma = \tol{1014}{+90}{-71}$ ~km~s$^{-1}$ and $\sigma =
  \tol{1032}{+99}{-76}$ ~km~s$^{-1}$ for A~222 and A~223,
  respectively.  While we find evidence for substructure in the
  spatial distribution of A~223, no kinematic substructure can be
  detected. From the red cluster sequence identified in a
  color--magnitude--diagram we determine the luminosity of both
  clusters and derive mass--to--light ratios in the $R$--band of
  $(M/L)_{\mathrm{A222}} = (202 \pm 43)~h_{70}~M_{\sun}/L_{\sun}$ and
  $(M/L)_{\mathrm{A223}} = (149 \pm 33)~h_{70}~M_{\sun}/L_{\sun}$.
  Additionally we identify a group of background galaxies at $z \sim
  0.242$.
} 
\maketitle
\section{Introduction}
\label{sec:introduction}
A~222/223 are two Abell clusters at $z \approx 0.21$ separated by
$\sim14\arcmin$ on the sky, or $\sim2600h_{70}^{-1}$ kpc, belonging to
the \citet{1983ApJS...52..183B} photometric sample. Both clusters are
rich having Abell richness class 3 \citep{1958ApJS....3..211A}. While
these are optically selected clusters, they have been observed by
ROSAT \citep{1997MNRAS.292..920W,1999ApJ...519..533D} and are
confirmed to be massive clusters.  9 spectra of galaxies in the
cluster region, most of them being cluster members, were known
\citep{1976ApJ...205..688S,1988ApJ...335..629N} before
\citet[][hereafter PEL]{2000A&A...355..443P} published a list of 53
spectra and did a first kinematical study of this system. PEL also
found 4 galaxies at the cluster redshift in the region between the
clusters (hereafter ``intercluster region''), indicating a possible
connection between the clusters.

We report 184 independent redshifts for 183 galaxies in the field of
Abell~222 and Abell~223, more than three times the number of redshifts
previously known, as well as equivalent widths for a number of lines.

The paper is organized as follows. In
Sect.~\ref{sec:data-data-reduction} we describe the reduction of the
spectroscopic and photometric data and discuss deviations from
previous values in the literature. The spatial distribution and the
kinematics of the double cluster system are examined in
Sect.~\ref{sec:spat-distr-kinem} with an emphasis on finding
possible substructure. We determine the luminosity and mass--to-light
ratio of the clusters by selecting the red cluster sequence in
Sect.~\ref{sec:mass-light-ratio}. Our results are summarized in
Sect.~\ref{sec:conclusions}. Throughout this paper we assume an
$\Omega_\Lambda = 0.7,\; \Omega_\mathrm{m} = 0.3,\;
H_0=70~h_{70}$~km~s$^{-1}$~Mpc$^{-1}$ cosmology.
\section{Data and Data Reduction}
\label{sec:data-data-reduction}
Multi-object spectroscopy of the two clusters Abell~222 and Abell~223
was performed at the NTT on three consecutive nights in December 1999.
These nights were clear with occasional high cirrus. The instrument
used was EMMI with grism 2, which has a resolution of 580 at 600~nm
and a dispersion of 11.6~nm/mm. With the 2048x2048 CCD pixels of
24~$\mu$m this leads to a dispersion of 0.28~nm/pixel. With one
exception two exposures of 2700 seconds each were taken for 6 fields,
3 on each cluster. For the field centered on A~222 in the second night
only one exposure of 2700 seconds was available. The wavelength
calibration was done using Helium--Argon lamps, which provided
typically 20 lines used in the calibration. The calibration frames
were taken at the beginning of the night for the masks used during
that night, before the science exposures were made.

\subsection{Reduction of spectroscopic data}
\label{sec:reduct-spectr-data}
For the data reduction a semi-automated IRAF\footnote{IRAF is
  distributed by the National Optical Astronomy Observatories, which
  are operated by the Association of Universities for Research in
  Astronomy, Inc., under cooperative agreement with the National
  Science Foundation.} package was written by the authors that cuts
out the single spectra of the CCD frames and then processes these
spectra using standard IRAF routines for single slit spectroscopy. The
sky spectrum was removed from all spectra using a linear fit with a $2
\sigma$ rejection on measurements on each side of the galaxy spectrum
where the position of the galaxy on the slit permitted it.
Measurements from only one side of the spectrum were used otherwise. A $2
\sigma$ rejection was used to remove cosmic rays and hot pixels.
Remaining hot pixels or cosmic rays in the sky spectrum introduced
fake absorption features, while hot pixels or cosmic rays in the
spectrum itself lead to fake emission features. These were removed by
hand.

Because the sky spectrum removal was done column by column and no
distortion correction was applied, some residual sky lines remained in
the final spectra, most notably of the strong [\ion{O}{i}] emission at
5577~\AA. At the typical redshift of the cluster members of $z \approx
0.21$ this line does not coincide with any important feature and thus
does not cause any problems in the subsequent analysis.

\subsection{Redshift determination}
\label{sec:redsh-determ}
The radial velocity determination was carried out using the 
cross-correlation method \citep{1979AJ.....84.1511T} implemented in
the RVSAO package \citep{1998PASP..110..934K}. Spectra of late type
stars and elliptical galaxies with known radial velocities were used
as templates. The redshift determination was verified by visual
inspection of identified absorption and emission features.
\begin{figure}
  \resizebox{\hsize}{!}{\includegraphics{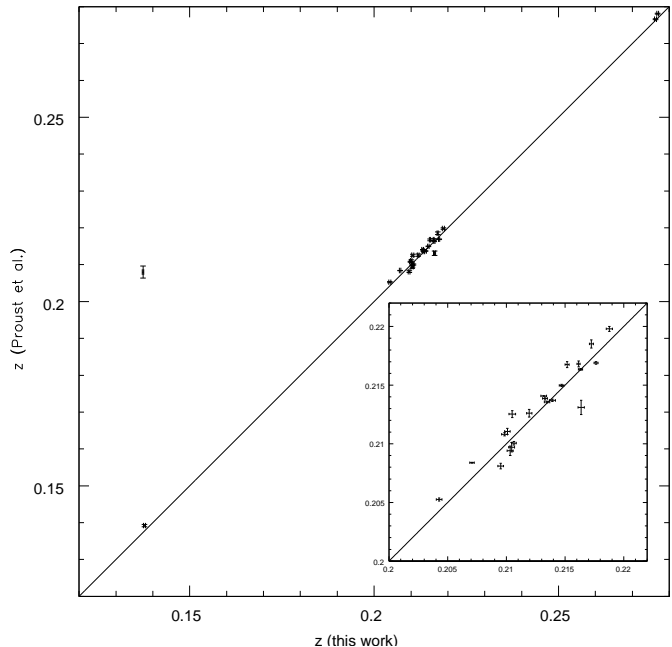}}
  \caption{Comparison of redshift measurements for objects observed by
    us and with redshifts listed in PEL. The large panel shows the
    full sample, the inset is a blow up of the cluster region. The
    error bars are the internal errors reported by RVSAO.}
  \label{fig:z-comp.all}
\end{figure}
The average internal error reported by RVSAO is $cz=68$~km~s$^{-1}$.
Adding this in quadrature to the estimated error of the wavelength
calibration of $cz=110$~km~s$^{-1}$, we estimate the error in
the redshift determination of single galaxies to be $\delta z=0.0004$.

Fig.~\ref{fig:z-comp.all} shows a comparison of our redshift
measurements and the redshifts listed by PEL. The obvious outlier in
the large panel is from the sample of \cite{1976ApJ...205..688S}.  The
inset shows a broad agreement between our results and the values of
PEL. Ignoring the obvious outlier, the average difference between the
measurements is $\left<z - z_\mathrm{PEL} \right> = (0.2\pm1.1) \times
10^{-3}$. Student's t-test rejects the null hypothesis of different
sample means with same variance at higher than the 99\% level for
the 23 cluster galaxies we have in common with PEL. However, Student's
t-test confirms the hypothesis of different sample means with same
variance at higher than the 99\% level for all their and our cluster
members, indicating that PEL observed a sub--sample with a
significantly different mean value from the larger sample we describe
here. 

We ruled out the possibility that this discrepancy could have been
caused by taking all calibration frames before the science exposures
were taken. This could have introduced a shift of the
zero point if the masks were not moved back to their original position
for the science exposures. We confirmed that this is not the case by 
determining the radial velocity of the subtracted sky spectrum in the
wavelength calibrated frames. We found that, if a zero point shift
occurred, it must be smaller than 30~km~s$^{-1}$, confirming the
accuracy of our data.

\subsection{Equivalent widths}
\label{sec:equivalent-widths}
We measured equivalent widths for the [\ion{O}{ii}]$\lambda$3727,
[\ion{O}{iii}]$\lambda$5007 emission lines, and H$\beta$ and H$\alpha$
emission and absorption lines. The integration ranges for the features
and the continuum were fixed by the values given in Table
\ref{tab:eqw-wavlen}.
\begin{table*}[htbp]
  \caption{Restframe wavelength ranges for equivalent widths measurement. All
    wavelengths are given in \AA. The last column gives the continuum
    range that was used for estimating the signal--to--noise ratio.} 
  \begin{tabular}{llllll}\hline
    Feature & $\lambda_{\mathrm{cent}}$ & line & blue cont. & red
    cont. & SNR \\\hline 
    $[\ion{O}{ii}]$ & 3727 & 3713 - 3741 & 3653 - 3713 & 3741 - 3801 &
    3560 - 3680 \\ 
    $[\ion{O}{iii}]$ & 5007 & 4997 - 5017 & 4872 - 4932 & 5050 - 5120 &
    4450 - 4750 \\ 
    H$\beta$ & 4861 & 4830 - 4890 & 4800 - 4830 & 4890 - 4920 & 4050 -
    4250 \\ 
    H$\alpha$ & 6563 & 6556 - 6570 & 6400 - 6470 & - & 6300 - 6450
    \\\hline  
  \end{tabular}
  \label{tab:eqw-wavlen}
\end{table*}

[\ion{O}{ii}] and H$\alpha$ are important indicators of star formation
rates \citep{1998ARA&A..36..189K}. To accurately determine the
equivalent widths and in particular estimate their significance level
we follow the definition of equivalent widths given by
\citet{2001A&A...372..391C}:
\begin{eqnarray}
  \label{eq:1}
  W_\lambda = \sum_{i=1}^{N_\mathrm{int}}
  \frac{f_i}{\overline{f_\mathrm{c}}}\Delta\lambda -
  N_\mathrm{int}\Delta\lambda,
\end{eqnarray}
where $f_i$ is the flux in pixel $i$, $N_\mathrm{int}$ is the number
of pixels in the integration range, $\overline{f_\mathrm{c}}$ is the
continuum level estimated as the mean of the continuum regions on
either side of the line, and $\Delta\lambda$ is the dispersion in
\AA/pixel. Note that with this definition emission lines have
\emph{positive} equivalent widths. 

The significance of an equivalent width measurement is given by
\citep{2001A&A...372..391C} 
\begin{eqnarray}
  \label{eq:2}
  \lefteqn{\sigma^2_{W_\lambda} = } \nonumber \\
  & & \left(\frac{S}{N}\right)^{-2} \left[ \left(
  W_\lambda + N_\mathrm{int} \Delta\lambda\right)\Delta\lambda +
  \frac{\left( W_\lambda + N_\mathrm{int}
  \Delta\lambda\right)^2}{N_\mathrm{c}} \right],
\end{eqnarray}
where the signal to noise ratio $S/N = \overline{f_\mathrm{c}} /
\sigma_\mathrm{c}$, and $\overline{f_\mathrm{c}}$ is obtained by
averaging over $N_\mathrm{c}$ pixel in the wavelength range given in
Table \ref{tab:eqw-wavlen}.

All wavelengths are given in the restframe of the object. All spectra
were normalized to a continuum fit before equivalent widths were
measured. The catalog lists all [\ion{O}{ii}] and [\ion{O}{iii}]
emission features and all H$\beta$ and H$\alpha$ emission and
absorption features that were detected with a significance $>2\sigma$. 

\subsection{Photometry}
\label{sec:photometry}
Wide-field imaging of the cluster pair was performed over two nights
in December 1999 with the Wide Field Imager on the ESO/MPG 2.2m on La
Silla.  Eleven 900 second exposures in $R$-band and three 900 second
exposures in $V$ band were taken using a dithering pattern which
filled the gaps between the CCDs in the mosaic in the coadded image.
The image reduction was carried out using a combination of
self--written routines and routines which are part of the IMCAT
software package written by Nick Kaiser ({\tt
  http://www.ifa.hawaii.edu/$\sim$kaiser/imcat}).  The images were
flattened with medianed night-sky flatfields from all the $R$ or $V$
band long exposure images taken over the two nights.  The images were
aligned using a process which assumes each CCD in the mosaic can be
translated to a common detector--plane coordinate system using a linear
transformation of coordinates (a shift in both axes and rotation
allowed) and that the detector-plane coordinates can then be
transformed into sky coordinates using a two dimensional polynomial,
in this case a bi-cubic polynomial.  The linear transformation from
each CCD to the detector-plane is assumed to be constant for all the
images whereas the transformation from detector-plane to sky
coordinates is determined separately for each image to allow for both
the pointing offsets in the dithering pattern and any changes in the
distortion pattern between images.  By comparing the positions of
stars among the individual images and to the positions in the USNO
catalog, both systems of equations for coordinate transformations were
solved using $\chi ^2$ minimization of the final stellar positions.
The rms dispersion of the centroids of the stars used in the fitting
were $0\farcs 016$ among the input images and $0\farcs 54$ between the
input images and the USNO coordinates, with the average offset vector
being consistent with zero in all regions of the image.  Further
details of this technique along with justifications for the linear
translation between CCD and detector plane can be found in
\citet{2001A&A...379..384C}.  The mapping of each input CCD was
performed using a triangular method with linear interpolation which
preserves surface brightness even if the mapping changes the area of a
pixel. The resulting images were then averaged using a 3$\sigma$
clipping algorithm to remove cosmic rays and moving objects. The final
$R$-band image can be found in Fig.~\ref{fig:den}.

\begin{figure}[htbp]
  \resizebox{\hsize}{!}{\includegraphics{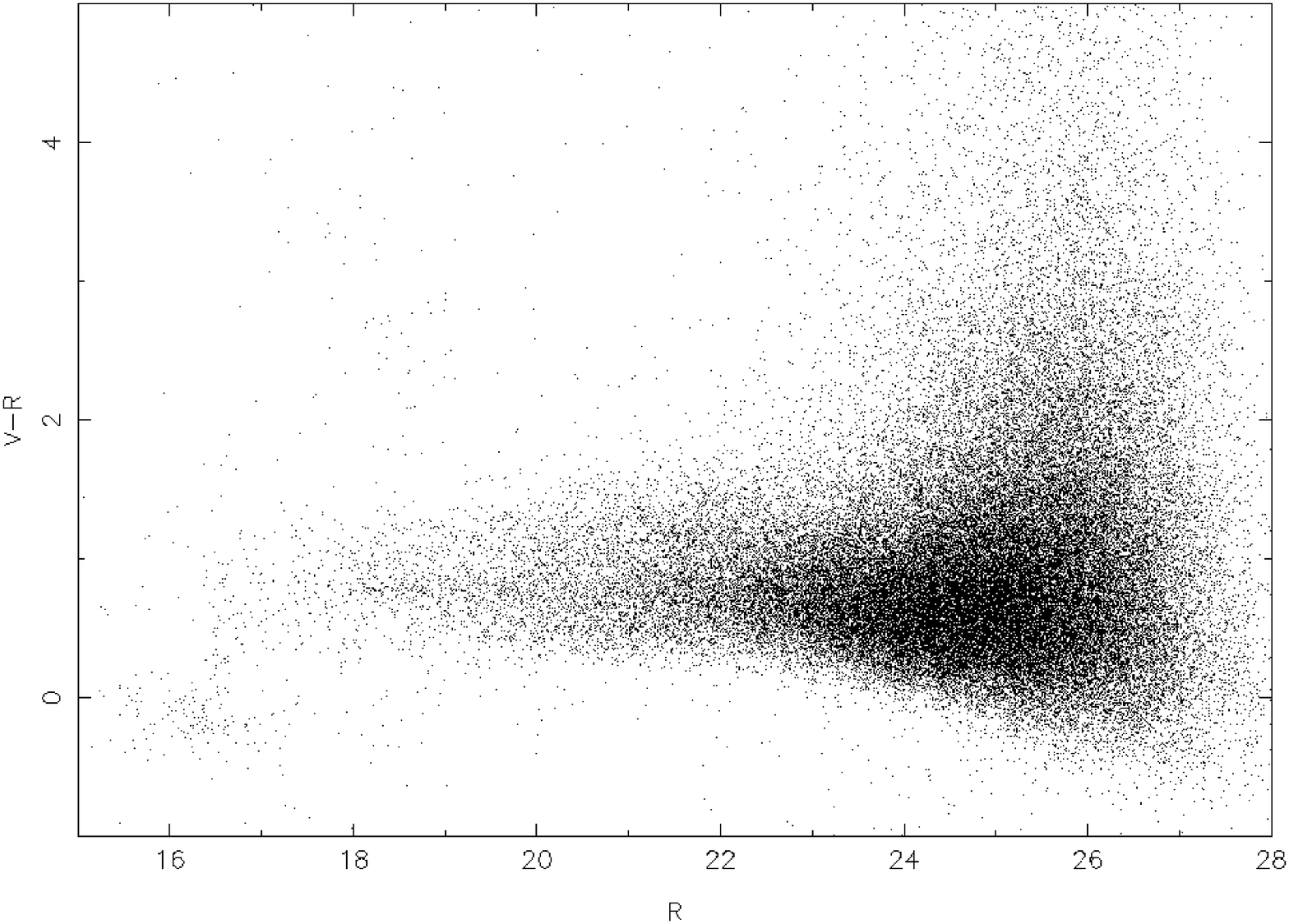}}
  \caption{Color--magnitude plot with the already corrected $V$
    magnitudes. The red cluster sequence is centered around \mbox{$V-R=
    0.8$}.} 
  \label{fig:cmd}
\end{figure}

Objects were detected in the $R$-band image using SExtractor
\citep{1996A&AS..117..393B}, and the $V$-band magnitudes for the
objects were measured using SExtractor in two-image mode.  The FWHM of
bright but unsaturated stars in the coadded images are $0\farcs 87$
for $R$ and $1\farcs 05$ in $V$.  Zeropoints were measured from
Landolt standard fields \citep{1992AJ....104..340L}, but the $V$-band
data is known to have been taken in non-photometric conditions.  From
isolating the red cluster galaxy sequence in a color-magnitude plot
(Fig.~\ref{fig:cmd}), corrected for the $A_B = 0.086$ mag dust
extinction \citep{1998ApJ...500..525S} using the conversion factors
from \citet{1989ApJ...345..245C}, and comparing to predicted colors of
cluster elliptical galaxies in a passive evolution model
\citep{1995PASP..107..945F}, a correction of $-0.23$ mag has been
applied to the $V$ magnitudes to correct for the additional
atmospheric extinction.  This correction also causes the stellar $V-R$
colors to have the theoretically expected values
\citep{1983ApJS...52..121G}.
All magnitudes are isophotal magnitudes with the limiting isophote at
27.96 mag/$\mathrm{arcsec}^2$. We determine the completeness limit of the
photometric catalog to be at $R=24$~mag from the point where the
number counts of objects departs from a power law.

An excerpt from the big catalog of spectroscopically measured galaxies
listing the 10 brightest galaxies in $R$--band in both cluster regions
can be found in Table~\ref{cat-exc}.  The full catalog is available in
electronic form at the at the Centre de Donn\'ees astronomiques de
Strasbourg (CDS)\footnote{\tt http://cdsweb.u-strasbg.fr/}.
\begin{sidewaystable*}
  \caption{Excerpt from the full catalog listing the 10 brightest
    galaxies in $R$--band in both cluster regions. The column entries
    are 1. Object No. 2. right ascension (hour:minute:second) 3.
    declination (degree:minute:second) 4. $R$ band magnitude 5.  $V-R$
    color 6--9.  observed equivalent widths of [\ion{O}{ii}],
    [\ion{O}{iii}], H$\beta$, and H$\alpha$ 10. \& 11.  heliocentric
    redshift and internal error reported by RVSAO 12. R-value of
    \protect{\citet{1979AJ.....84.1511T}} 13.  notes, {\bf p}:
    \protect{\citet{2000A&A...355..443P}}, {\bf n}:
    \protect{\citet{1988ApJ...335..629N}}, {\bf s}:
    \protect{\citet{1976ApJ...205..688S}}, {\bf em}: radial velocity
    derived from emission line template.}
  \begin{tabular}[h]{rrrrrccccrrrl}
\hline\noalign{\smallskip}
Object No.&RA (2000)&Dec
(2000)&$R$&$V-R$&[\ion{O}{ii}]/\AA&[\ion{O}{iii}]/\AA&H$\beta$/\AA&H$\alpha$/\AA&$z_\mathrm{hel}$&$\sigma_{z_\mathrm{hel}}$&
R&Notes\\
\noalign{\smallskip}
\hline\hline
\noalign{\smallskip}
{\bf ABELL 222 }&&&&&&&&&&&&\\
93&01:37:41.54&$-$12:58:30.8&16.91&0.89&-&-&$-4.0\pm1.6$&-&0.21318&0.00024&4.14&\\
&&&&&&&&&0.21409&0.00022&&p \\
30&01:37:26.34&$-$12:59:56.9&16.97&0.83&-&-&$-7.0\pm1.4$&-&0.21347&0.00023&5.34&\\
&&&&&&&&&0.21356&0.00007&&p \\
65&01:37:34.01&$-$12:59:28.6&17.00&0.89&-&-&$-7.7\pm1.6$&-&0.21329&0.00021&6.11&\\
&&&&&&&&&0.21387&0.00012&&p \\
8&01:37:17.97&$-$13:01:20.7&17.03&0.83&-&-&$-7.5\pm1.7$&-&0.21132&0.00011&11.30&\\
97&01:37:43.03&$-$12:57:43.7&17.41&0.84&-&-&$-6.7\pm1.6$&-&0.21392&0.00026&4.86&\\
&&&&&&&&&0.21371&0.00010&&n \\
69&01:37:34.2&$-$12:56:52.7&17.44&1.01&-&-&$-6.8\pm1.9$&-&0.20809&0.00028&4.05&\\
16&01:37:22.65&$-$13:00:21.3&17.60&0.76&-&-&$4.9\pm1.4$&-&0.21161&0.00019&6.13&\\
13&01:37:21.21&$-$13:00:39.2&17.74&0.81&-&-&$-5.9\pm1.3$&-&0.22067&0.00020&5.97&\\
39&01:37:28.32&$-$12:55:55.7&17.77&0.66&-&$15.7\pm0.5$&$9.3\pm1.2$&$42.6\pm0.7$&0.05147&0.00023&4.35&$0.05126\pm0.00013$
measured on [\ion{O}{ii}], \\
&&&&&&&&&&&&[\ion{O}{iii}], H$\beta$, H$\alpha$ \\
95&01:37:42.08&$-$12:55:35.5&17.96&0.75&-&$1.9\pm0.9$&$-3.4\pm1.3$&-&0.21612&0.00019&5.87&\\
\dots & \dots & \dots & \dots & \dots & \dots & \dots & \dots & \dots
& \dots & \dots & \dots &\\
{\bf ABELL 223}&&&& &&&&&&&&\\
173&01:38:02.30&$-$12:45:19.5&16.58&0.83&$4.6\pm1.9$&$3.2\pm2.0$&$-3.5\pm1.6$&-&0.20427&0.00025&4.55&$0.20459\pm0.00009$
measured on[\ion{O}{iii}],\\ 
&&&&&&&&&&&&[\ion{N}{ii}]\\
&&&&&&&&&0.20525&0.00013&&p \\
&&&&&&&&&0.20506&0.00031&&p \\
&&&&&&&&&0.20600&0.00167&&s \\
186&01:38:04.83&$-$12:47:34.0&16.88&0.83&-&-&$-4.7\pm2.1$&-&0.20972&0.00025&5.07&\\
167&01:38:01.02&$-$12:46:51.6&17.00&0.79&-&-&$-4.6\pm1.6$&-&0.21118&0.00024&5.14&\\
140&01:37:56.02&$-$12:49:09.8&17.12&0.84&-&-&$-6.7\pm1.9$&-&0.21050&0.00028&4.34&\\
&&&&&&&&&0.21253&0.00030&&p \\
172&01:38:02.19&$-$12:45:41.1&17.34&0.91&-&-&$-7.3\pm1.7$&$-0.8\pm0.3$&0.24074&0.00023&5.68&\\
119&01:37:50.32&$-$12:46:15.4&17.36&0.75&-&-&$-3.5\pm1.4$&-&0.20132&0.00018&6.66&\\
193&01:38:07.25&$-$12:48:13.0&17.36&0.87&-&-&$-4.4\pm1.5$&-&0.20957&0.00028&4.44&\\
149&01:37:57.22&$-$12:50:17.1&17.40&0.68&$36.9\pm3.4$&$15.8\pm0.8$&-&$14.0\pm2.4$&0.20707&0.00019&6.67&$0.20694\pm0.00038$
measured on [\ion{O}{ii}], \\
&&&&&&&&&&&&[\ion{O}{iii}], H$\beta$ \\
&&&&&&&&&0.20839&0.00004&&p \\
154&01:37:57.74&$-$12:47:55.8&17.41&0.83&-&-&$-7.7\pm1.7$&-&0.21045&0.00025&4.82&\\
&&&&&&&&&0.20971&0.00040&&n \\
198&01:38:15.49&$-$12:43:38.4&17.48&0.80&-&-&$-5.2\pm1.2$&$3.9\pm0.9$&0.21560&0.00023&4.69&\\
\dots & \dots & \dots & \dots & \dots & \dots & \dots & \dots & \dots
& \dots & \dots & \dots &\\
\noalign{\smallskip}\hline
\end{tabular}

  \label{cat-exc}
\end{sidewaystable*}

\section{Spatial Distribution and Kinematics}
\label{sec:spat-distr-kinem}
After removing some obvious background and foreground galaxies
($z>0.3$ or $z<0.1$), an iterative $3 \sigma$ clipping was used to
decide upon cluster membership. We found 81 galaxies belonging to
Abell~222 and 72 galaxies belonging to Abell~223 or the possible
bridge connecting both clusters.

\begin{figure}
  \resizebox{\hsize}{!}{\includegraphics{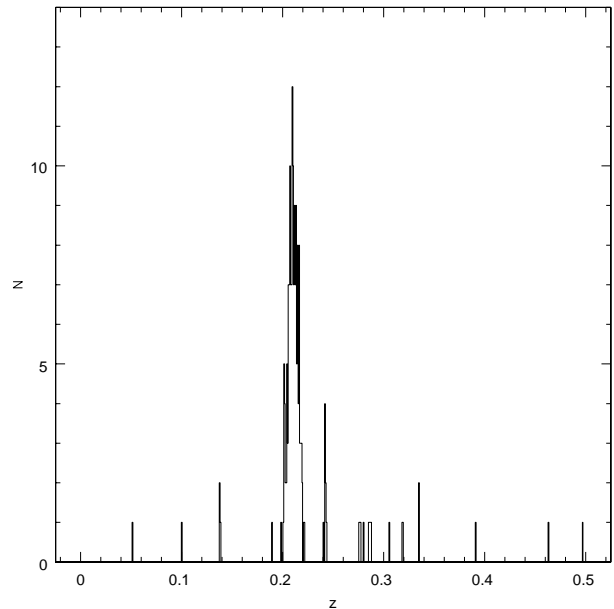}}
  \caption{Radial velocity distribution of all galaxies in the
    sample. The main peak corresponds to the two Abell clusters.}
  \label{fig:ALL_hist}
\end{figure}

\begin{figure}
  \resizebox{\hsize}{!}{\includegraphics{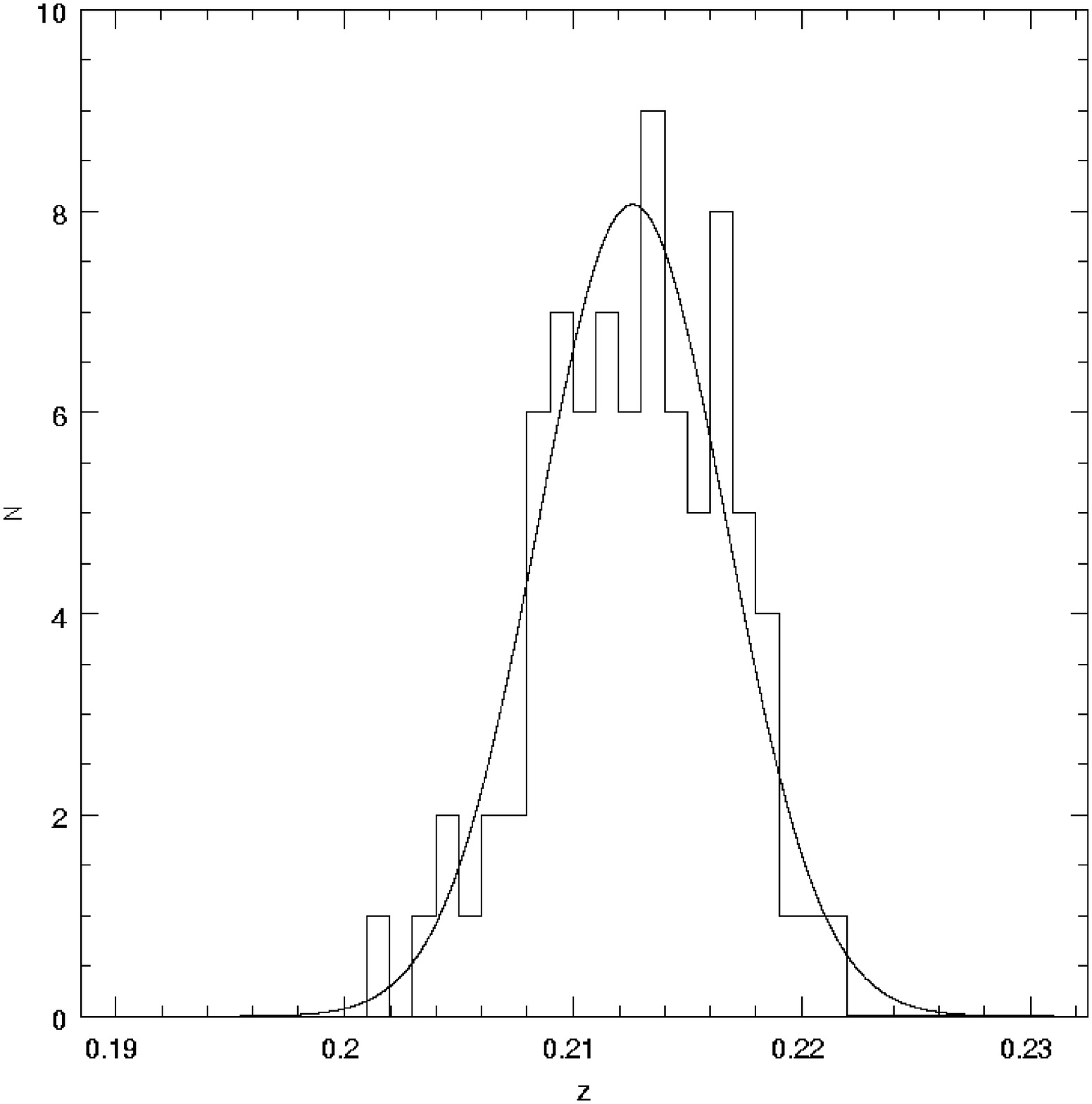}}
  \caption{Radial velocity distribution of the members of A~222. The
    continuous line is a Gaussian with mean and measured dispersion
    value as given in the text.}
  \label{fig:A222_hist}
\end{figure}

\begin{figure}
  \resizebox{\hsize}{!}{\includegraphics{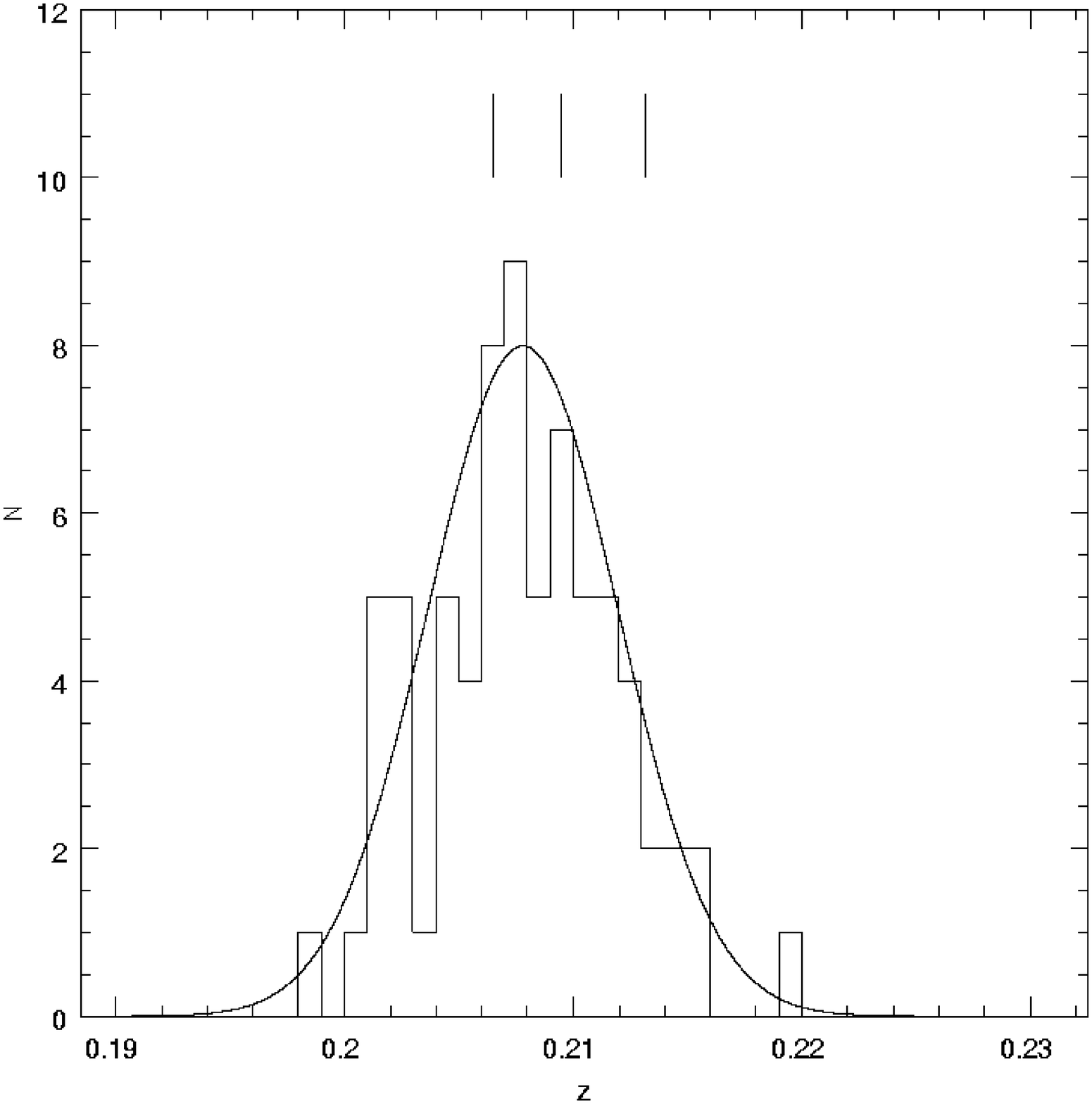}}
  \caption{Same as Fig.~\ref{fig:A222_hist} for A~223.
    The dashes denote the 3 galaxies in the bridge for which we
    measured radial velocities.}
  \label{fig:A223_hist}
\end{figure}

The mean redshift of the individual clusters are $z =0.2126 \pm
0.0008$ and $z = 0.2079 \pm 0.0008$, for A~222 and A~223,
respectively. This differs significantly from the values of $z=0.2143$
and $z=0.2108$ found by PEL for A~222 and A~223, respectively. The
quoted errors include the statistical error of $\delta z = 0.0005$
added linearly to the estimated error of the cluster member selection
of $\delta z = 0.0003$.  This systematic error was estimated from
varying the cut level of the recursive clipping procedure from
$2\sigma$ to $3.5\sigma$ and calculating the means of these cuts. The
statistical errors were calculated from a bootstrap resampling of the
cluster members. The measured velocity dispersions have to be
transformed to the restframe of the cluster according to the
transformation law
\begin{eqnarray}
  \label{eq:9}
  \sigma_\mathrm{cor} = \frac{\sigma}{1+\overline{z}}~,
\end{eqnarray}
\citep{1974ApJ...191L..51H}. The restframe velocity dispersions for
the individual clusters are $\sigma_\mathrm{cor} =
\tol{1014}{+90}{-71}$~km~s$^{-1}$ and $\sigma_\mathrm{cor} =
\tol{1032}{+99}{-76}$~km~s$^{-1}$, for A~222 and A~223, respectively.
This is in good agreement with the values found by PEL of $\sigma =
1013\pm150$~km~s$^{-1}$ and $\sigma=1058\pm160$~km~s$^{-1}$ for A~222
and A223, respectively. The errors of the velocity dispersion are the
$(1-\alpha)=68\%$ confidence interval given by
\begin{eqnarray}
  \label{eq:10}
  \frac{(n-1)\sigma^2}{\chi^2_{n-1;\alpha/2}} < \sigma^2 <
    \frac{(n-1)\sigma^2}{\chi^2_{n-1;1-\alpha/2}}~, 
\end{eqnarray}
for $n$ galaxies in the sample. The redshift and velocity dispersion
for A~223 do not change significantly if the 3 galaxies from the
intercluster region in our sample are removed.
Figs.~\ref{fig:ALL_hist} to~\ref{fig:A223_hist} show the corresponding
radial velocity distributions of the individual samples.


The values for the velocity dispersion are somewhat higher than
those derived from X--ray luminosities. \citet{1999ApJ...519..533D}
report bolometric luminosities of $L_\mathrm{X} = 7.65 \times
10^{44}$~erg~s$^{-1}$ and $L_\mathrm{X} = 6.94 \times
10^{44}$~erg~s$^{-1}$ from ROSAT PSPC observations for A~222 and
A~223, respectively, for $H_0=50$~km~s$^{-1}$~Mpc$^{-1}$. Using the
$L_\mathrm{X}$--$\sigma$ relationships of \citet{1999ApJ...524...22W}
we get $\sigma_\mathrm{X} = 845-887$~km~s$^{-1}$ for A~222 and
$\sigma_\mathrm{X} = 828-871$~km~s$^{-1}$ for A~223. Compared to the
velocity dispersions derived from the spectroscopic observations, both
cluster appear to be underluminous in X--rays.

Together with the data of PEL we now have radial velocities for 6
galaxies in the possible bridge connecting both clusters. With our new
values for the radial velocity of A~223 the observation made by PEL,
that most of the bridge galaxies are in the low--velocity tail, does
not hold anymore. In fact they all appear to be close to the maximum
or higher of the velocity histogram shown in Fig.~\ref{fig:A223_hist}.
Because A~222 is the cluster at higher redshift, this is the expected
behavior should these galaxies indeed belong to a bridge connecting
both clusters

\begin{figure*}
  \centering
  \resizebox{\hsize}{!}{\includegraphics[angle=270]{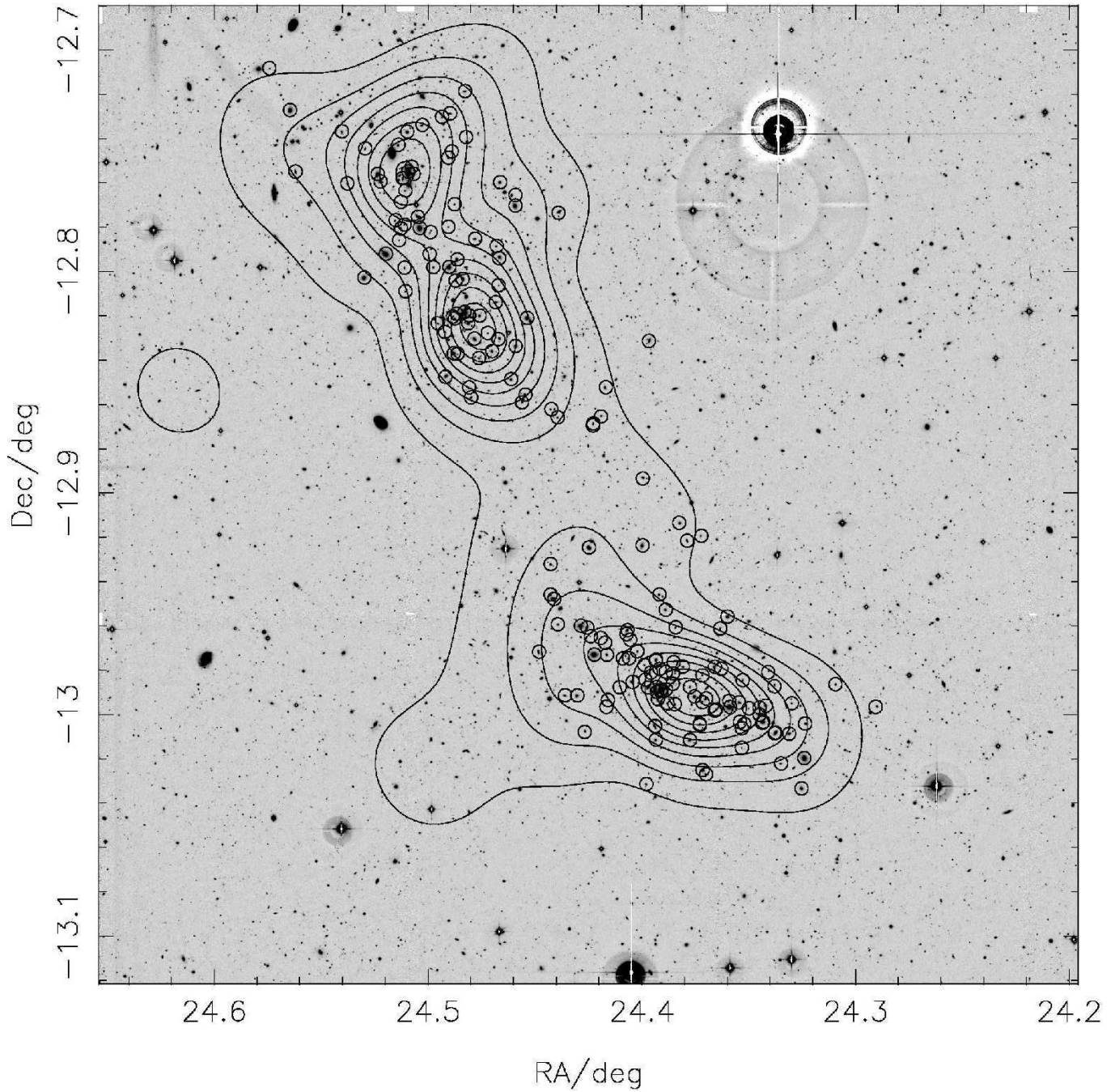}} 
  \caption{Deep $R$ band image of A~222 and A~223. The galaxy density
    contours are generated from a color selected sample of 702
    objects. See text for details. The small circles mark all
    spectroscopically identified cluster galaxies.}
  \label{fig:den} 
\end{figure*}

SExtrator also provides an algorithm to separate galaxies from stars.
Each object is assigned a CLASS\_STAR value, which is 1 for stars, 0
for galaxies, and lies in between for ambiguous objects.  Fig.
\ref{fig:den} displays a projected galaxy number density map generated
with the adaptive kernel density estimate method described by
\citet{1996MNRAS.278..697P} for a color selected sample of 702 objects
with $R < 21$ and $0.7 < V-R <0.9$ and SExtractor CLASS\_STAR $< 0.1$
from the WFI images.  Overplotted are the positions of all
spectroscopically identified cluster members.
Fig.~\ref{fig:den} clearly exhibits two density peaks in A~223. Both
peaks are separated by $4\farcm8$ and are centered at
($\alpha$=01:37:53.5, $\delta$=$-$12:49:21.2) and ($\alpha$=01:38:01.8,
$\delta$=$-$12:45:07.0). These peaks are also visible in the density
distribution of the 181 spectroscopically identified cluster galaxies.
Also visible is an overdensity of color selected objects in the
intercluster region, hinting at a connection between both clusters.

We applied the \citet[][DS]{1988AJ.....95..985D} test for local
kinematic deviations in the projected galaxy distribution. The DS test
is based on computing deviations of local mean velocity and velocity
dispersion from the global values. The local values are calculated for
each of the $N$ galaxies and its $n-1$ nearest neighbors.  The
statistics used to quantify the presence of substructure is
\begin{eqnarray}
  \label{eq:3}
  \Delta & = & \sum_{i=1}^{N} \delta_i \nonumber \\
  & = & \sum_{i=1}^{N}\left\{\frac{n}{\sigma_\mathrm{glob}^2}
      \left[
        \left(\overline{v}_\mathrm{glob} -
          \overline{v}_\mathrm{loc,i}\right)^2
        + \left(\sigma_\mathrm{glob} - \sigma_\mathrm{loc,i}\right)^2
      \right]\right\}^{\frac{1}{2}}
\end{eqnarray}
The $\Delta$ statistics is calibrated by randomly shuffling the radial
velocities while keeping the observed galaxy positions fixed. The
significance of the observed $\Delta_\mathrm{obs}$ is assessed by the
fraction of simulations whose $\Delta_\mathrm{sim}$ is smaller than
$\Delta_\mathrm{obs}$.

We find that, while the DS test is clearly able to separate both
clusters at better than the 99.9\% confidence level, it does not find
any substructure in the individual clusters for values of $8<n<16$.

Also the DIP statistic \citep{1985ANNSTAT.13..70H}, which we calculated
with the FORTRAN routine provided by \citet{1985APPLSTAT.34..320}, is
not able to reject the null hypothesis of an unimodal distribution for
any of the cluster samples. We tested for deviations from a normal
Gaussian distribution by computing the skewness and kurtosis of both
cluster samples. We found that we cannot reject a Gaussian parent
population for both cluster samples at the $1 \sigma$ level.

\begin{figure}
  \resizebox{\hsize}{!}{\includegraphics{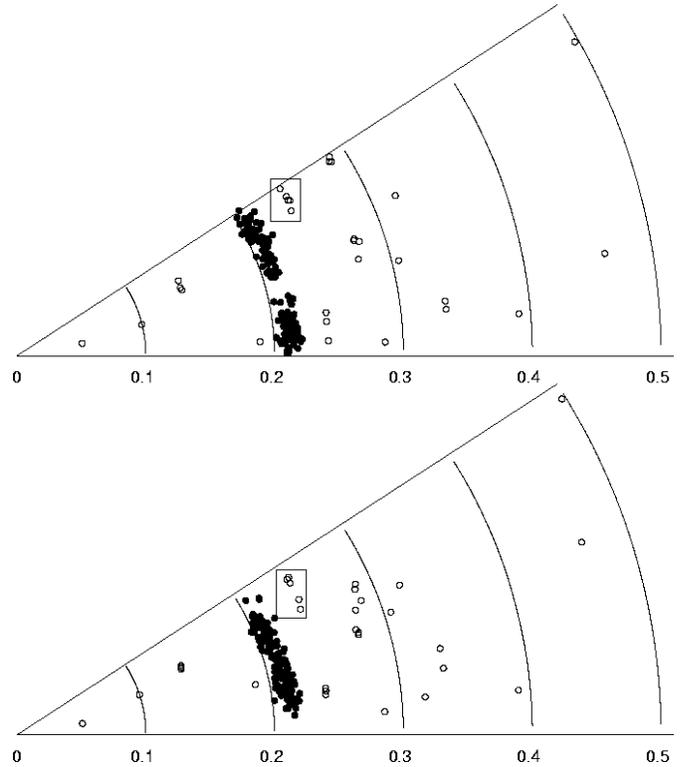}}
  \caption{Declination (top) and right ascension (bottom) wedge
    redshift diagram. Cluster members are plotted as solid circles
    while background and foreground galaxies are displayed with open
    symbols. The rectangle highlights a small background group of
    galaxies. The opening angle is from $-$13\degr 2\arcmin
    24\arcsec~to $-$12\degr 16\arcmin 24\arcsec~for the declination
    wedge and from 01:37:04.8 to 01:38:48.0 for the right ascension
    wedge.}
  \label{fig:wedges2}
\end{figure}
The wedge velocity diagrams in Fig.~\ref{fig:wedges2} clearly show the
Abell system at $z \sim 0.21$. A small group of five galaxies can be
seen behind A~223 at $z = 0.242$. We derive a velocity dispersion of
$\sigma = 330$~km~s$^{-1}$, confirming that it is not only close in
the projected spatial distribution but also in redshift space.

\section{Mass--to--light ratio}
\label{sec:mass-light-ratio}
To determine the luminosity of the clusters we applied the same cut on
color and CLASS\_STAR as above in a circle with 1.4~$h_{70}^{-1}$~Mpc
radius around the bright cD galaxy of A~222 and the center of the line
connecting both density peaks in A~223. We binned the selected objects
in bins of 0.5 mag and fitted a \citet{1976ApJ...203..297S} luminosity
function,
\begin{eqnarray}
  \label{eq:4}
  n(L)\mathrm{d}L = n^*(L/L^*)^\alpha \exp(-L/L^*)\mathrm{d}(L/L^*)~,
\end{eqnarray}
where $L^*$ is the characteristic luminosity, $\alpha$ is the
faint--end slope, and $n^*$ is a normalization constant. Written in
terms of absolute magnitude Eq.~(\ref{eq:4}) becomes
\begin{eqnarray}
  \label{eq:5}
  N(M)\mathrm{d}M = k n^* \exp\left\{\left[
    -k(\alpha+1)(M-M^*)\right] \right. \nonumber \\
  - \left. \exp\left[-k(M-M^*)\right]\right\}\mathrm{d} M~,
\end{eqnarray}
where $M^*$ is the absolute magnitude corresponding to $L^*$ and $k =
\ln10/2.5$ \citep{1995ApJ...452L..99K}. Using k correction and a
passive evolution correction on the synthetic elliptical galaxy
spectra of \citet{1993ApJ...405..538B}, for the Abell system $M_R =
m_R -5 \log\left(\frac{d_\mathrm{l}}{\mathrm{Mpc}}\right) - 24.91$,
$d_\mathrm{l}$ being the luminosity distance.

The fit is performed by minimizing the quantity
\begin{eqnarray}
  \label{eq:6}
  \chi^2 = \sum \frac{\left[N(M_i) - N_f(M_i)\right]^2}{\sigma_i^2}~,
\end{eqnarray}
with $N(M_i)$ and $N_f(M_i)$ being the observed and fitted number of
galaxies in the $i$th magnitude bin. The variance of galaxies in each
magnitude bin was assumed to be that of a Poissonian distribution.

\begin{figure}
  \resizebox{\hsize}{!}{\includegraphics{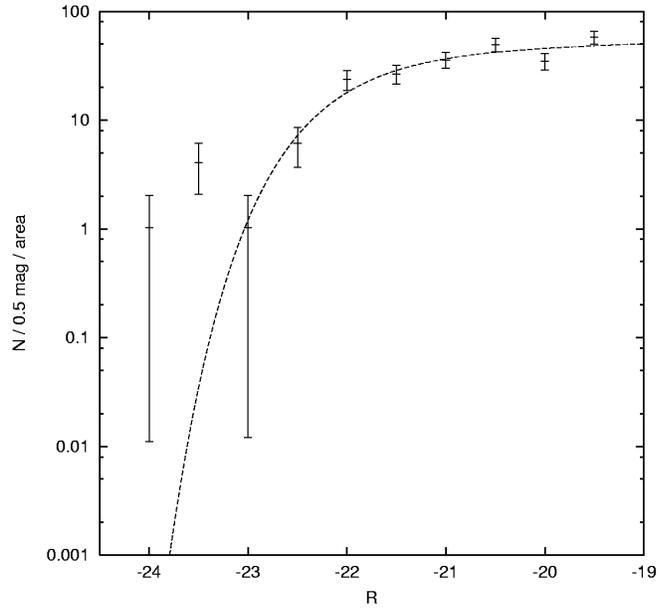}}  
  \caption{Differential $R$--band luminosity function for A~222. The
    points represent the objects selected by the criteria detailed in
    the text, while the dashed line is the best--fit Schechter
    function.} 
  \label{fig:schechter222}
\end{figure}

\begin{figure}
  \resizebox{\hsize}{!}{\includegraphics{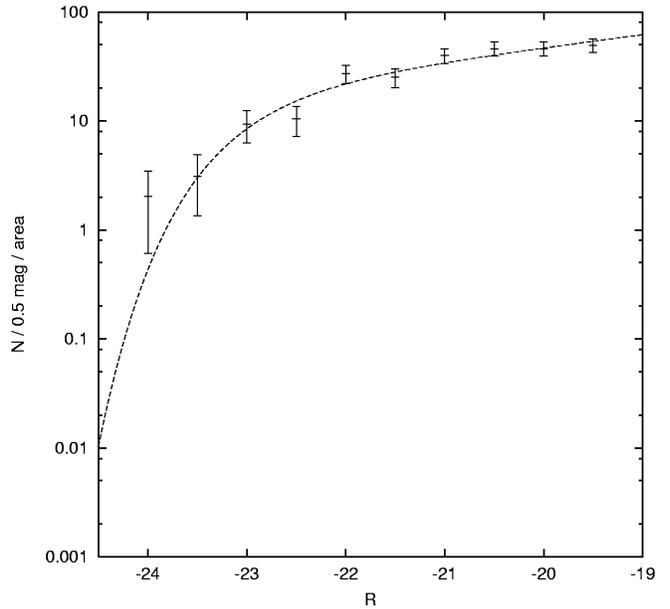}}
  \caption{Same as Fig.~\ref{fig:schechter222} for A~223.}
  \label{fig:schechter223}
\end{figure}

The best--fit Schechter function for A~222 has $M^* = -22.1 \pm 0.2$,
$\alpha = -1.04 \pm 0.12$, and $n^* = 31 \pm 11$. The $\chi^2$ value
for these parameters is 12.7 with 7 degrees of freedom. The best--fit
parameters for A~223 are $M^* = -23.1\pm0.2$, $\alpha=-1.20\pm0.06$,
and $n^* = 15\pm 5$ with a minimum $\chi^2$ value of 7.0, also for 7
degrees of freedom. From Figs.~\ref{fig:schechter222} and
\ref{fig:schechter223} we see, that the Schechter function is a good
representation of the faint end, while it slightly underpredicts the
number of bright galaxies. Although the values of $M^*$ differ by one
magnitude they are compatible with values from the literature. E.g.
\citet{1996A&A...315..365T} find $M^*=-22.66\pm0.52$ from a study of
36 Abell clusters for a fixed $\alpha = -1.25$. For this value of
$\alpha$ the $M^*$ value for A~222 increases to $-22.4$, but the fit
then has a reduced $\chi^2$ of 2.2. In the following we only use the
lower value of $M^*=-22.1$ for A~222.

The total $R$--band luminosity of the red cluster sequence is
determined by extrapolating the luminosity function. The observed
fraction of the total luminosity is given by $\Gamma(\alpha+2,
L_\mathrm{lim}/L^*)/\Gamma(\alpha+2)$, where $\Gamma(\cdot, \cdot)$ is
the incomplete Gamma function, $L_\mathrm{lim}$ is the completeness
limit, which in this case is given by the selection parameters, and
$L^*$ is the luminosity corresponding to the fitted $M^*$
\citep{2001AJ....122.1289T}.  It follows from the chosen magnitude cut
and the size of the bins, that in our case the limiting magnitude $M_R
= -19.0$. This implies that we observe $91.4\%$ and $81.7\%$ of the
total light in A~222 and A~223, respectively. The total $R$--band
luminosity of the red cluster sequence then is $L_R = (3.1 \pm 0.3)
\times 10^{12}\;h_{70}^{-2}\; L_{\sun}$ for A~222 and $L_R = (4.4 \pm
0.4) \times 10^{12} \;h_{70}^{-2}\;L_{\sun}$ for A~223, where the
errors reflect the uncertainty in the parameters of the luminosity
function.

We compute mass--to--light ratios assuming an isothermal sphere model
for both clusters with the velocity dispersion determined in
Sect.~\ref{sec:spat-distr-kinem}. The mass of an isothermal sphere
inside a radius $r$ is given by
\begin{eqnarray}
  \label{eq:7}
  M(<r) = \frac{\pi \sigma_\mathrm{cor}^2}{G}r~.
\end{eqnarray}
We get the following mass--to--light ratios inside $1.4~h_{70}^{-1}$~Mpc:
\begin{eqnarray*}
  \label{eq:8}
  \mathrm{A~222}:\;(M/L)_R & = & (343 \pm 60)~h_{70}~M_{\sun}/L_{\sun}\;,\\
  \mathrm{A~223}:\;(M/L)_R & = & (253 \pm 47)~h_{70}~M_{\sun}/L_{\sun}\;.
\end{eqnarray*}
By selecting only the red cluster sequence we miss a significant part
of the cluster luminosity. \citet{1999MNRAS.308..459F} found a ratio
between the luminosity of their ``type 1'' galaxies (E/S0) and the
total luminosity of $0.59\pm0.07$ in the fields of the 2dF survey.
Using this correction factor we arrive at the final result:
\begin{eqnarray*}
  \mathrm{A~222}:\;(M/L)_R & = & (202 \pm 43)~h_{70}~M_{\sun}/L_{\sun}\;,\\
  \mathrm{A~223}:\;(M/L)_R & = & (149 \pm 33)~h_{70}~M_{\sun}/L_{\sun}\;.
\end{eqnarray*}
These values slightly overestimate the true mass--to--light ratio because 
the use of isophotal magnitudes cuts off part of the galaxy luminosity
and the intracluster light.

If we use the velocity dispersion derived from X--ray measurement
instead of the spectroscopically determined velocity dispersion we
arrive at mass--to--light ratios that are lower by $\sim 20\%$ but
still agree within the $1\sigma$ error with the values quoted above.

\section{Conclusions}
\label{sec:conclusions}
We have reported 184 independent redshifts measurements for 183
galaxies in the field of Abell~222 and Abell~223, as well as
equivalent widths for [\ion{O}{ii}], [\ion{O}{iii}], H$\beta$, and
H$\alpha$, $R$ magnitudes, and $V-R$ color.

From a sample of 153 galaxies which we identified as cluster members,
we derived a mean redshift and restframe velocity dispersion of $z
=0.2126 \pm 0.0008$, $\sigma_\mathrm{cor} =
\tol{1014}{+90}{-71}$~km~s$^{-1}$ for A~222 and $z = 0.2078 \pm
0.0008$, $\sigma_\mathrm{cor} = \tol{1032}{+99}{-76}$~km~s$^{-1}$ for
A~223. The values of the redshifts are clearly outside the error
margins of the values previously reported by PEL.  By comparing our
wavelength calibration to the sky spectrum, which provides an
independent wavelength standard, we were able to confirm the accuracy
of our data and rule out the possibility of a zero point shift of more
than 30~km~s$^{-1}$ for each mask. $R$ and $V$ band photometry was
taken from WFI data.

Although the projected density maps of all spectroscopically
identified galaxies and of a color selected sample with 702 members
clearly show spatial substructure in A~223, neither the DS test nor
the DIP statistics were able to find any kinematic substructure. Also
no indications of a non--Gaussian parent population could be found.

We fitted a Schechter luminosity function to objects in the red
cluster sequence identified in a color--magnitude diagram. Assuming an
isothermal sphere model for the cluster we derived $(M/L)$ ratios in
$R$--band, which are comparable for both clusters. The computed values
are $(M/L)_R = (202\pm43)~h_{70}~M_{\sun}/L_{\sun}\;$ and $(M/L)_R =
(149\pm33)~h_{70}~M_{\sun}/L_{\sun}$ for A~222 and A~223,
respectively.  This is within the range of values reported by other
groups for other cluster.

\citet{1978ApJ...226...55D} gave a range of
$140-420~h_{70}~M_{\sun}/L_{\sun}$ in a study of 12 rich clusters.
Typical values for virial mass--to--light ratio are at values of $M/L
\sim 210~h_{70}~M_{\sun}/L_{\sun}$ \citep{1996ApJ...462...32C}.
Typical values derived from X--ray masses tend to be somewhat lower
than those from virial masses. \citet{2000ApJ...543..521H} find a
median value of $(M/L)_V \sim 140~h_{70}~M_{\sun}/L_{\sun}$ in a study
of eight nearby clusters and groups.

We cannot exclude the possibility that the values we report here are
biased towards higher values by using an isothermal sphere model. Both
cluster geometries clearly deviate from circular symmetric profiles.
More robust mass estimates may thus lead to lower $(M/L)$ ratios.

A detailled discussion whether the galaxies between both clusters
indeed belong to a structure connecting the cluster pair will be part
of a forthcoming weak lensing study of this system.

\begin{acknowledgements}
  We thank Joan--Marc Miralles and Lindsay King for help and useful
  discussions. This work was supported by the TMR Network
  "Gravitational Lensing: New Constraints on Cosmology and the
  Distribution of Dark Matter'' of the EC under contract No.
  ERBFMRX-CT97-0172.
\end{acknowledgements}

\bibliographystyle{aa}
\bibliography{h3727}

\end{document}